\documentclass{elsarticle}
\usepackage{amsmath}
\usepackage{lineno,hyperref}
\modulolinenumbers[5]

\journal{Astroparticle Physics}

\begin{document}

\begin{frontmatter}

\title{On the Estimation of the Depth of Maximum of Extensive Air Showers Using the Steepness Parameter of the Lateral Distribution of Cherenkov Radiation }

\author{Ibrahim Rasekh, Davoud Purmohammad\fnref{myfootnote}}
\address{Physics Department, Imam Khomeini International University, Qazvin, Iran}
\fntext[myfootnote]{Corresponding author}




\begin{abstract}
  Using Monte Carlo simulation of extensive air showers, we showed that the maximum depth of showers, $X_{max}$ can be estimated using $P=Q(100)/Q(200)$, the ratio of Cherenkov photon densities at 100 and 200 meters from the shower core, which is known as the steepness parameter of the lateral distribution of Cherenkov radiation on the ground.  A simple quadratic model has been fitted to a set of data from simulated extensive air showers, relating the steepness parameter and the shower maximum depth. Then the model has been tested on another set of simulated showers. The average difference between the actual maximum depth of the simulated showers and the maximum depth obtained from the lateral distribution of Cherenkov light is about 9 g/cm\textsuperscript{2}.  In addition, possibility of a more direct estimation of the mass of the initial particle from $P$ has been investigated. An exponential relation between these two quantities has been fitted. Applying the model to another set of showers, we found that the average difference between the estimated and the actual mass of primary particles is less than 0.5 atomic mass unit.
\end{abstract}


\end{frontmatter}


\section{Introduction}
Extremely energetic cosmic rays are accessible signs of phenomena such as supernova explosions, or active galactic nuclei in very distant parts of space. Determining characteristics such as energy and mass of cosmic rays helps us to identify their origin. Because of the very low flux of cosmic rays in very high energy, it is almost impossible to directly observe them outside the atmosphere. On the other hand, these energetic particles produce extensive showers of secondary charged particles in the atmosphere. The detection of these secondary charged particles, atmospheric fluorescence radiation, Cherenkov radiation, or radio waves emanating from the showers, are the main methods of detecting extensive air showers (EAS). Characteristics such as the size of a shower and the atmospheric depth at which the shower reaches its maximum size, which depend on the energy and type of the initial particle generating the shower, can be estimated by arrays of detectors on the ground. Studies on the structure of extensive air showers have shown that the maximum depth of showers  correlate with their energy and primary mass \cite{Bellido17,Berezhnev12,Corstanje21,Chernov05,Takeishi18,Arbeletche20}. Relatively direct measurement of the maximum depth of a shower is only possible with atmospheric fluorescence detector arrays which measure the amount of isotropic radiation from the passage of shower particles through the atmosphere, along the shower path \cite{Abbasi18}. On the other hand, particle detector arrays, or Cherenkov radiation detectors on the ground, can only observe the distribution of particles or Cherenkov radiation on a single observation level, i.e., the lateral distributions, and are not able to inspect the distribution of particles or radiation along the shower axis. Research has shown the correlation between the lateral distribution of Cherenkov radiation in an extensive air shower and the height of
the shower maximum \cite{Lubsandorzhiev08}. The height can then be converted to the atmospheric depth  by using an atmospheric density distribution model, e.g. section 6.1 of the reference \cite{Stanev10}. Hence, the depth of the shower maximum can be estimated from the lateral distribution of its Cherenkov radiation. By analyzing the results form a large number of simulated extensive air showers, here we will establish an empirical relationship between the depth of the shower maximum and the steepness parameter of its Cherenkov radiation lateral distribution, with no need to the intermediate   step of the height estimation. This provides a more accurate method for the estimation of the depth of shower maximum from the Cherenkov radiation lateral distribution.  We  also show that the mass of the primary particle can be estimated by using an empirical model that establishes an immediate relationship between the steepness parameter of the Cherenkov radiation distribution and the mass of the primary particle.

\section{Simulation}
\begin{figure}
\centering\includegraphics[width=0.8\textwidth]{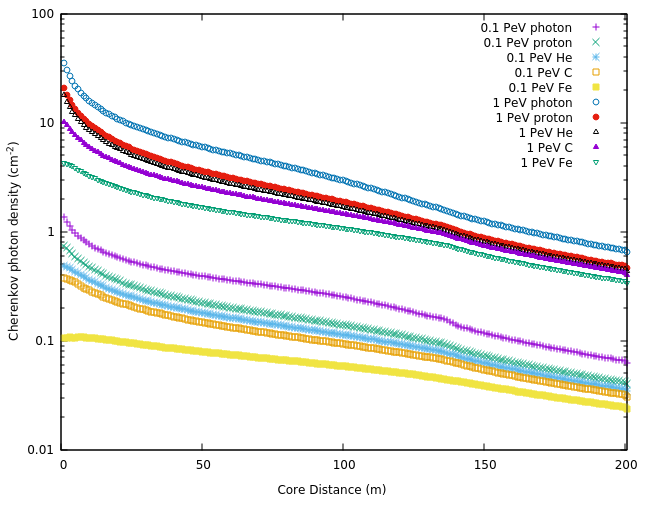}
\caption{  An example of lateral distribution of Cherenkov radiation for $10^{14}$ eV and $10^{15}$ eV showers initiated by gamma ray (photon), proton, helium, carbon and iron primaries, simulated with QGSJETII+GHISHA models for an observation level 675 meter above sea level. }\label{fig1}
 \end{figure}

  9890 extensive air showers with energies from $10^{12}$ to $10^{17}$ eV, 
initiated by different primary particles, such as gamma ray, proton,
iron, carbon, helium, silicon, oxygen, magnesium, aluminum, sulfur and
scandium were produced by using CORSIKA simulation code \cite{Heck98}. Most of these showers, which have been used for the establishment of the proposed methods, were generated by using QGSJETII model \cite{Ostapchenko06} for high energy, and
GHEISHA model \cite{Fesefeldt85} for low energy hadronic interactions. In order to test the impact of the hadronic interaction models on the viability of our proposed methods, we used EPOS-LHC  model \cite{Pierog15} for high energy and UrQMD model \cite{Bass98} for low energy hadronic interactions to generate some other showers.   The number of the simulated showers for different primary types and energyies for each part of the analysis is given in the relevant sections. No thinning has been applied. The simulations were set to generate longitudinal profiles of the particles in steps of 10 g/cm\textsuperscript{2}, and fit the profile to the Gaisser-Hillas type distribution specified in the CORSIKA documents \cite{Heck98}.  The $t_{max}$ parameter of the Gaisser-Hillas distribution has been used as an estimation of the depth of maximum ($X_{max}$) of a shower. However, we have  observed that, in some showers, the difference between evaluated $t_{max}$ and the depth of the step corresponding to the maximum number of charged particles in the longitudinal output of CORSIKA is greater than the step size, which is 10 g/cm\textsuperscript{2}. In these cases, the sampled depth of maximum charged particles was taken as  $X_{max}$. The showers were simulated for the observation level
of 675 meters above the sea level, which is the height of the
Tunka-133 Cherenkov array \cite{Lubsandorzhiev08}. This array is one of the few
experiments in which the lateral distribution of Cherenkov radiation of
showers is measured. The absorption of Cherenkov radiation in the
atmosphere and the default quantum efficiency for detectors have been
taken into account.  No photon bunch size has been specified in the simulations inputs; hence, the default values were implemented ( CERSIZ=0.).  By inspecting the data in the output files, we observed that the maximum number of Cherenkov photons in a bunch was 30.   The detector array responses were
not simulated in this work. The showers produced in the simulations had
zero zenith angle, and their energies were in the $10^{12}$ eV to $10^{17}$ eV interval. For
shower energies less than $10^{16}$ eV, all Cherenkov
photons within 201meter distance from shower cores have been simulated.
 An array of $81\times 81$ flat detectors, with single detector area of 1
m\textsuperscript{2}, and 5 m inter detector spacing has been
implemented to reduce simulation time and output file size for higher energy showers. The outputs of a CORSIKA shower simulation contain
information including the number and the location of the Cherenkov
photons reaching the observation level, and the number of secondary
particles in 10 g/cm\textsuperscript{2} steps from the top to the bottom
of the atmosphere. Therefore, the lateral distribution of Cherenkov
radiation at the observation level and the depth of shower maximum have
been calculated from the simulation results for each shower. To
determine the lateral distribution of Cherenkov radiation of a shower,
we divided the observation level into annuli of one meter in width around the shower core. The number of photons entered into each annulus was used to
determine the number of photons per unit area, $Q(r)$ in each
region of outer radius $r$ from the core. Then, the steepness of the lateral distribution  $P=Q(100)/Q(200)$,  in which $Q(100)$ and $Q(200)$ are photon densities at 100 and 200 meter form core respectively, has been calculated. An example of lateral distribution of Cherenkov radiation for $10^{14}$ eV and $10^{15}$ eV showers initiated by gamma ray, proton, helium, carbon and iron primaries is shown in Figure \ref{fig1}. 
 
\section{ Estimation of the maximum depth of extensive air
showers using \emph{P}}
\begin{figure}
\centering\includegraphics[width=0.6\textwidth]{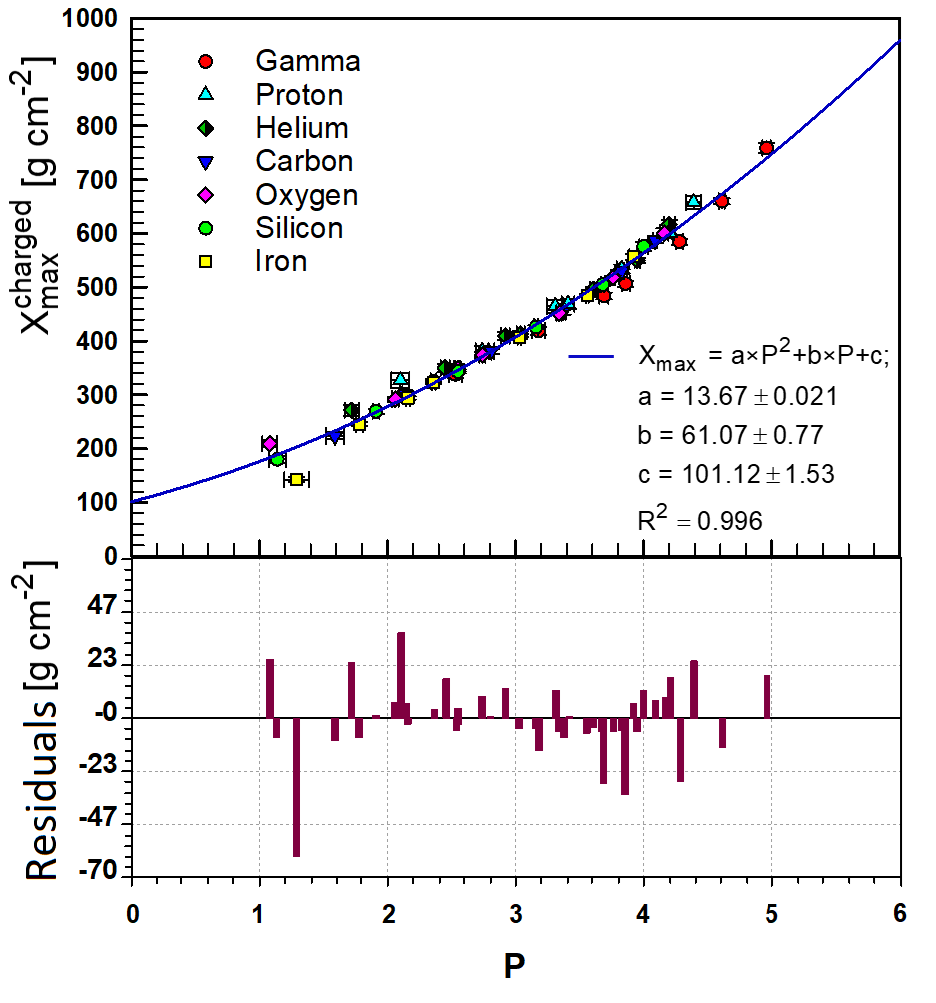}
\caption{Above: Variation of shower maximum with the steepness
 of the Cherenkov radiation lateral distribution along with a quadratic model for 4340
simulated showers, generated from different primary particles type and energies. The data points represent average values for showers of the same type-energy.  The number of showers for each type-energy is given in Table \ref{tabfig2}. The fit paremeters, and the coefficient of determination are also given.  Below: The residual plot for the fit. The horizontal scales are the same for both plots. }\label{fig2}
 \end{figure}

\begin{figure}
\centering\includegraphics[width=\textwidth]{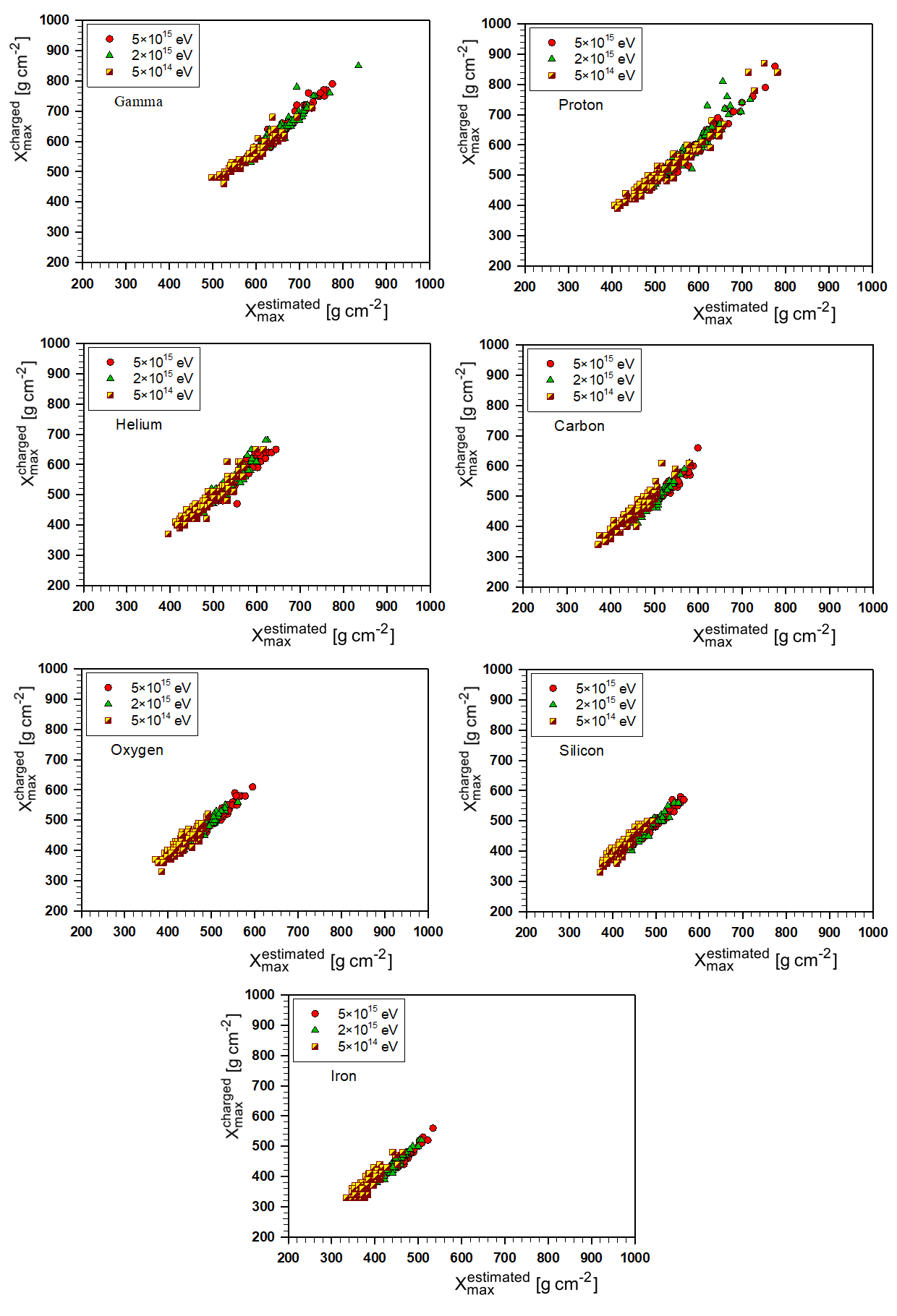}
\caption{  Comparison of shower maximum depths obtained from the
longitudinal distribution of charged particles and those estimated by
using equation \ref{equation1} for simulated showers initiated by  gamma rays, protons and
the nuclei of helium, carbon, oxygen, silicon, and  iron primary particles with
$5\times10^{14}$ eV, $2\times10^{15}$ eV, and $5\times10^{15}$ eV energies. The number of simulated showers used in these figures is given in Table \ref{tabfig3} .}\label{fig3}
 \end{figure}

\begin{figure}
\centering\includegraphics[width=0.5\textwidth]{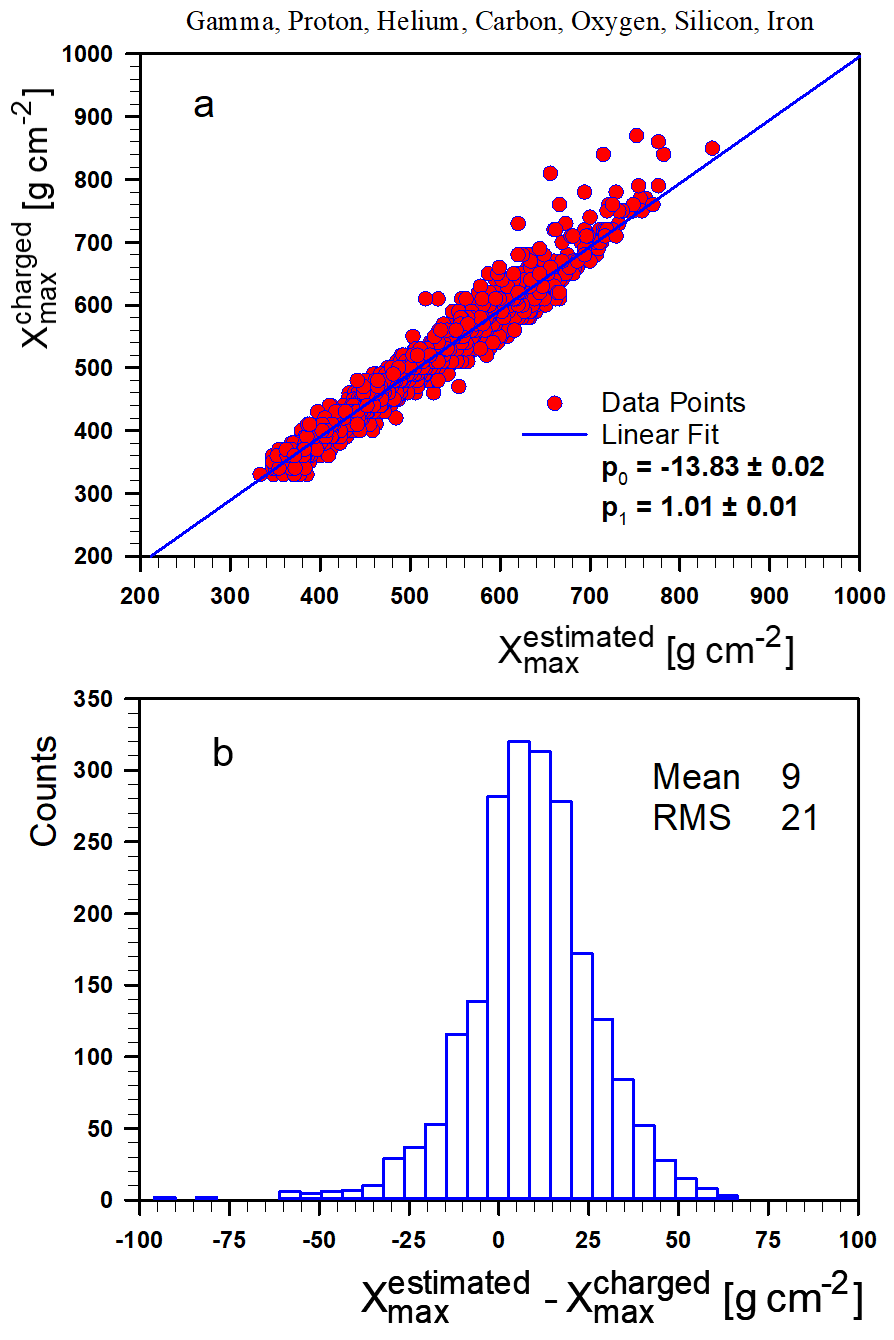}
\caption{ \emph{a}) The actual maximum depths obtained from longitudinal
distribution of charged particles have been compared with the maximum
depths estimated by equation \ref{equation1}, for a set of 2100 simulated showers
initiated by different primary particles, including  gamma rays, proton, He,
C, O,  Si and Fe with $5\times10^{14}$ eV,
$2\times10^{15}$ eV, $5\times10^{15}$ eV energies. The data points have been fitted to $X_{max}^{charged}=p_0+p_1 X_{max}^{estimated}$
. \emph{b}) Histogram of the deviation of the estimated maximum depths
from the actual ones, for the same data.}\label{fig4}
 \end{figure}

Since the shower maximum depth is the main quantity used for the estimation
of the primary particle masses, it would be interesting to remove the
intermediate step of maximum height estimation and find an immediate
relation between \emph{X\textsubscript{max}}, and the steepness of the
lateral distribution of Cherenkov radiation. Having tested several
models, we chose a quadratic model to relate
\emph{X\textsubscript{max}} and \emph{P}, for its simplicity and
suitable fit to the simulated data:
\begin{equation}
X_{max}=a P^2 + b P + c \label{equation1}
\end{equation}
 The model is fitted to the data of 4340 simulated extensive air showers of different primarey particle types and energies. The number of the simulated showers for each type-energy used in the fit is given in Table \ref{tabfig2}.   The fit and its parameters 
 are presented in Figure \ref{fig2}. 
\begin{table}[t]\caption{The number of simulated events used for fitting the data presented in Figure \ref{fig2} to equation \ref{equation1}.}
\centering
\resizebox{\textwidth}{!}{\begin{tabular}{|c|c|c|c|c|c|c|c|}
\hline
 & \multicolumn{7}{c|}{No. of showers} \\
\hline
Type$\backslash$ Energy & 10$^{12}$eV  & 10$^{13}$eV & 5x10$^{13}$eV & 
10$^{14}$eV & 10$^{15}$eV &10$^{16}$eV &10$^{17}$eV \\
\hline
$\gamma$ & 100 & 100 & 100 & 100 & 100 & 100 & 20 \\
\hline
 \it{p}  & 100 & 100 & 100 & 100 & 100 & 100 &20\\
\hline
 He  & 100 & 100 & 100 & 100 & 100 & 100 & 20 \\
\hline
C & 100 & 100 & 100 & 100 & 100 & 100 & 20 \\
\hline
O  & 100 & 100 & 100 & 100 & 100 & 100 & 20 \\
\hline
Si & 100 & 100 & 100 & 100 & 100 & 100 & 20 \\
\hline
Fe & 100 & 100 & 100 & 100 & 100 & 100 &  20\\
\hline
\end{tabular}\label{tabfig2}}\end{table}
 To test the model for $X_{max}$ estimation in single showers, we tried it using another
set of simulated showers as follows. The steepness parameter \emph{P} of each shower in the new set
was obtained and the estimated depth of maximum, $X_{max}^{estimated}$ was calculated by equation \ref{equation1}. The actual depth of maximum, $X_{max}^{charged}$ has also been obtained from the
longitudinal charged particle distribution. The comparison of the two
depths, for different primary masses and energies, has been represented
in Figure \ref{fig3}.  The number of simulated showers of different type-energies in Figure \ref{fig3} is given in Table \ref{tabfig3}.  The average values of $P$, $X_{max}^{estimated}$,  $X_{max}^{charged}$, and their relative difference for the data represented in Figure \ref{fig3} are given in  Table \ref{table2}. The correlation between the estimated and the actual maxima of all showers in Figure \ref{fig3}
has been shown in Figure \ref{fig4}-{\it a}. A histogram of the deviation of the estimated maxima is presented in Figure \ref{fig4}-{\it b}. As Figure \ref{fig4}-{\it b} shows, the average and RMS
of the differences are 9 g/cm\textsuperscript{2}, and 21
g/cm\textsuperscript{2}, respectively. 
\begin{table}[h]\caption{ Number of simulated events for testing equation \ref{equation1} for $X_{max}$ estimation of single showers in Figure \ref{fig3}.}
\centering
\resizebox{0.5\textwidth}{!}{\begin{tabular}{|c|c|c|c|}
\hline
 & \multicolumn{3}{c|}{No. of showers} \\
\hline
Type$\backslash$ Energy & 5x10$^{14}$eV  & 2x10$^{15}$eV & 5x10$^{15}$eV  \\
\hline
$\gamma$ & 100 & 100 & 100 \\
\hline
 \it{p}  & 100 & 100 & 100 \\
\hline
 He  & 100 & 100 & 100  \\
\hline
C & 100 & 100 & 100 \\
\hline
O  & 100 & 100 & 100 \\
\hline
Si & 100 & 100 & 100  \\
\hline
Fe & 100 & 100 & 100\\
\hline
\end{tabular}\label{tabfig3}}\end{table}
 
\begin{table}[ht] \caption{ The average values for the steepness of the lateral
distribution of Cherenkov radiation, \emph{P}, the maximum depths
obtained from longitudinal distribution of charged particles, $X_{max}^{charged}$ , the
maximum depths estimated by using equation \ref{equation1}, $X_{max}^{estimated}$, and the relative
difference of the two maxima, $\delta=\frac{\vert X_{max}^{charged} - X_{max}^{estimated}\vert}{X_{max}^{charged}}$, for the set of simulated extensive air
showers used in Figure \ref{fig3}. }
\centering
\resizebox{\textwidth}{!}{\begin{tabular}{|c|c|c|c|c|c|}
\hline
 Primary Particle & Primary Energy [eV] & $P=Q(100)/Q(200)$  & $X_{max}^{charged}$ [g/cm$^2$] & $X_{max}^{estimated}$ [g/cm$^2$] & $\delta$ \\
\hline
$\gamma$-ray  & $5\times10^{14}$  & $4.13 \pm0 .03$ & $553 \pm 5.1$ & $587 \pm 4.7$ &     0.061 \\              
        & $2\times10^{15}$         & $4.41 \pm 0.02$ & $613 \pm 5.9$ & $638 \pm 4.6$ &     0.041  \\  
        & $5\times10^{15}$         & $4.55 \pm 0.02$ & $643 \pm 5.7$ & $663 \pm 4.3$ &     0.031  \\  
\hline           
\it{p}  & $5\times10^{14}$  & $4.74 \pm 0.04$ & $519 \pm 9$ & $524 \pm 7.4$ &     0.010 \\              
        & $2\times10^{15}$         & $3.96 \pm 0.04$ & $557 \pm 8.2$ & $558 \pm 6.1$ &     0.002  \\             
        & $5\times10^{15}$         & $4.06 \pm 0.03$ & $574 \pm 7.5$ & $576 \pm 5.8$ &     0.003  \\             
\hline
He      & $5\times10^{14}$  & $3.49 \pm 0.03$ & $477 \pm 5.7$ & $482 \pm 4.5$ &     0.010 \\              
        & $2\times10^{15}$         & $3.72 \pm 0.02$ & $515 \pm 5.1$ & $519 \pm 3.9$ &     0.008  \\             
        & $5\times10^{15}$         & $3.92 \pm 0.02$ & $549 \pm 5.2$ & $551 \pm 3.9$ &     0.004  \\             
\hline
C       & $5\times10^{14}$  & $3.3 \pm 0.02$ & $448 \pm 5.2$ & $453 \pm 3.7$ &     0.011 \\              
        & $2\times10^{15}$         & $3.52 \pm 0.02$ & $476 \pm 4$ & $486 \pm 3.1$ &     0.021  \\             
        & $5\times10^{15}$         & $3.71 \pm 0.02$ & $509 \pm 4.1$ & $516 \pm 3.1$ &     0.014  \\             
\hline
O       & $5\times10^{14}$  & $3.18 \pm 0.02$ & $427 \pm 3.8$ & $434 \pm 2.9$ &     0.016 \\              
        & $2\times10^{15}$         & $3.47 \pm 0.02$ & $469 \pm 3.3$ & $478 \pm 2.6$ &     0.019  \\             
        & $5\times10^{15}$         & $3.66 \pm 0.02$ & $503 \pm 3.7$ & $509 \pm 2.9$ &     0.012  \\             
\hline
Si      & $5\times10^{14}$  & $3.08 \pm 0.02$ & $415 \pm 3.6$ & $419 \pm 2.5$ &     0.010  \\              
        & $2\times10^{15}$         & $3.34 \pm 0.02$ & $450 \pm 3.7$ & $458 \pm 3.1$ &     0.018  \\             
        & $5\times10^{15}$         & $3.57 \pm 0.02$ & $489 \pm 3.6$ & $494 \pm 2.8$ &     0.010  \\             
\hline
Fe      & $5\times10^{14}$  & $2.83 \pm 0.02$ & $380 \pm 3.3$ & $383 \pm 2.3$ &     0.008 \\              
        & $2\times10^{15}$         & $3.22 \pm 0.02$ & $431 \pm 3.1$ & $439 \pm 2.3$ &     0.019  \\      
        & $5\times10^{15}$         & $3.4 \pm 0.01$ & $458 \pm 2.7$ & $466 \pm 2.1$ &     0.017  \\      
        \hline       
\multicolumn{5}{|r|}{ $\langle\delta\rangle=$ } &0.016\\
\hline
\end{tabular}\label{table2}}\end{table}
\section{An investigation of immediate estimation of primary particles
mass using \emph{P}}
\begin{figure}
\centering\includegraphics[width=0.8\textwidth]{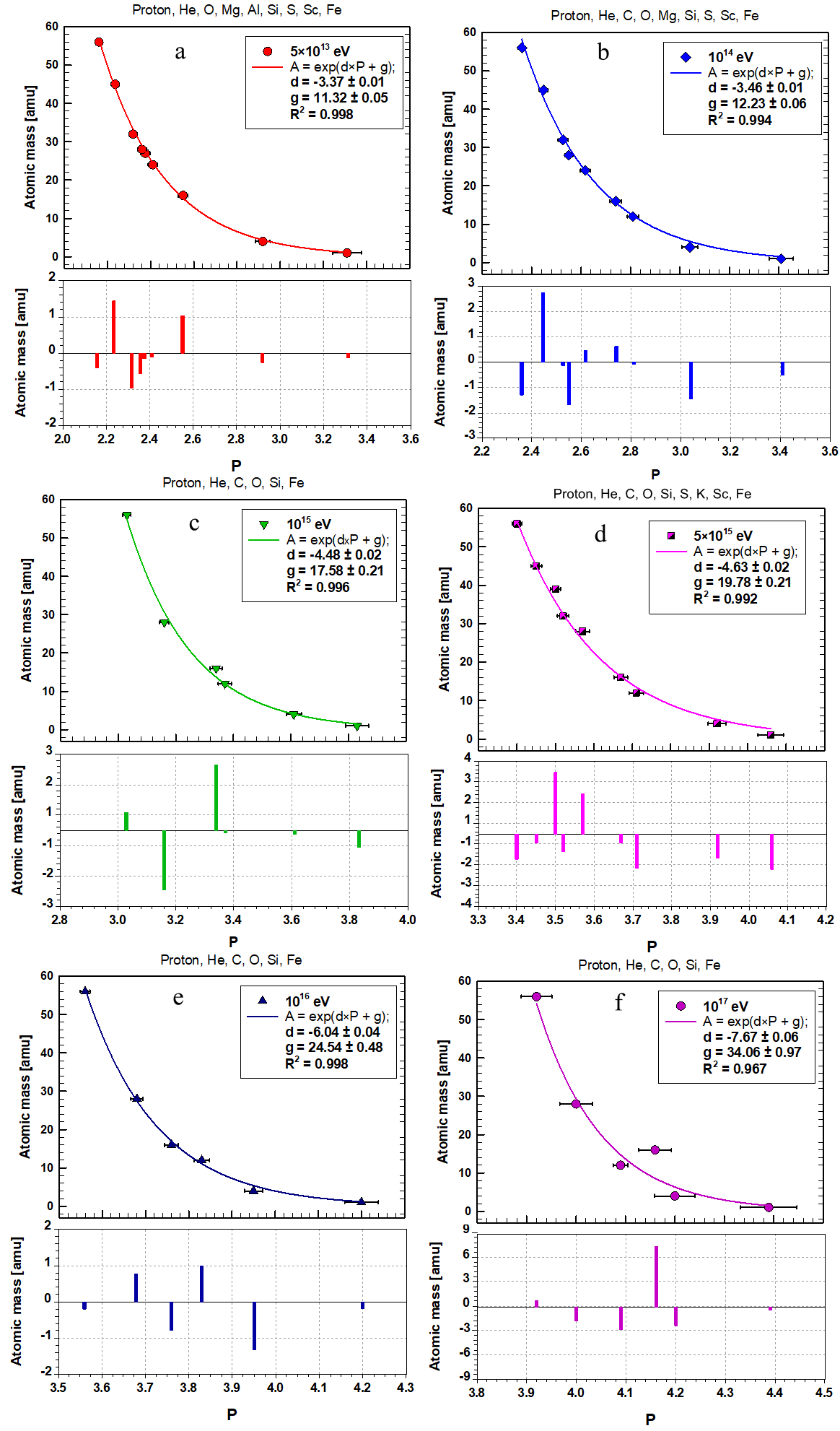}
\caption{ The results of fitting the equation \ref{equation2} to a set of simulated
extensive air showers initiated by different primary cosmic rays with
$5\times10^{13}$ eV, $10^{14}$ eV,
$10^{15}$ eV, $5\times10^{15}$ eV,
$10^{16}$ eV, and $10^{17}$ eV energies. Each
point is an average value among 100 showers, except for the
$10^{17}$ eV points which are average for 20 showers.  The fit residuals are plotted below each figure. The number of simulated showers used in these plots is given in Table \ref{tabfig5}.}\label{fig5}
 \end{figure}

\begin{figure}
\centering\includegraphics[width=0.5\textwidth]{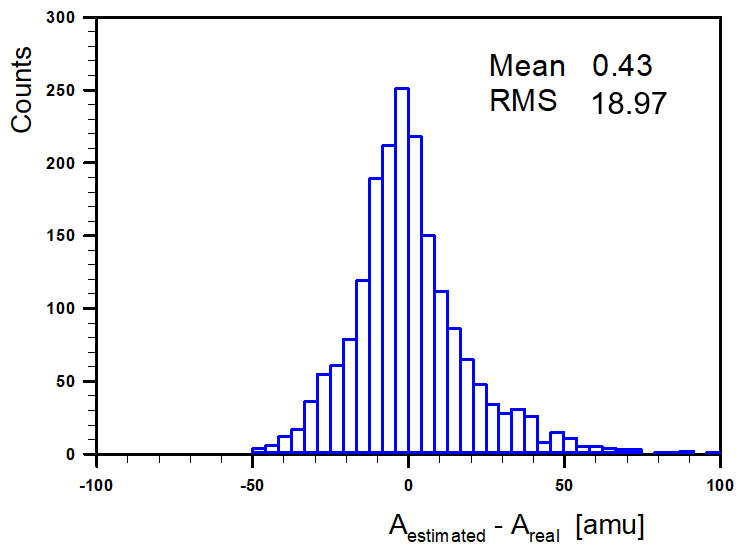}
\caption{ The distribution of differences between the actual masses of
the primary particles and the estimated masses obtained by application
of equation \ref{equation5}. The results are based on event-by-event evaluation for
1900 simulated extensive air showers initiated by primary masses from
\emph{A}=1 to \emph{A}=56 and with energies from
$10^{14}$ eV to $2\times10^{15}$ eV. The number of simulated showers for each type-energy is given in Table \ref{tabfig6}.}\label{fig6}
 \end{figure}

The main use of knowing the \emph{X}\textsubscript{max} of a shower is
for the estimation of the mass of its primary particle. If one can find a
model to relate the lateral distribution of Cherenkov radiation of an
EASs to its primary mass, there will be no need to estimate
\emph{X}\textsubscript{max}. Therefore, we tried to find a model
relating the shower primary particle mass to its steepness of the
lateral Cherenkov radiation distribution, for a set of simulated
extensive air showers. After the trial of a few models, we chose an
exponential relation with two parameters, for its simplicity and
suitable fit to the data:
\begin{equation}
A=\exp (d \times P + g), \label{equation2}
\end{equation}
where \emph{A} is the mass of the primary particle in atomic mass unit
(a.m.u.), and \emph{P} is the steepness parameter. The model is only
applicable to showers initiated by cosmic rays. The results of fitting
the model to the simulated extensive air showers of different primary
masses and energies have been shown in Figure \ref{fig5}.
\begin{table}[h]\caption{Number of simulated showers used  in Figure \ref{fig5} for fitting  equation \ref{equation2}.}
\centering
\resizebox{0.8\textwidth}{!}{\begin{tabular}{|c|c|c|c|c|c|c|c|c|c|c|c|c|}
\hline
 Energies[eV]  & p & He & C & O & Si & Fe & Mg & Al & S & K & Sc & Sum\\
\hline
5x10$^{13}$ & 100 &100 & & 100 & 100 & 100 & 100 & 100 &100 &  & 100 & 900  \\
\hline
10$^{14}$ & 100 &100 & 100 &100 & 100 & 100 & 100 &  & 100 &  & 100 & 900 \\
\hline
10$^{15}$  & 100 &100 &100 &100 &100 & 100 &  &  &  &  &  & 600 \\
\hline
5x10$^{15}$  & 100 & 100 & 100 &100 &100 & 100 &  &  & 100 & 100 & 100 & 900 \\
\hline
10$^{16}$  &100 &100 &100 &100 &100 & 100 &  &  &  &  &  & 600 \\
\hline
10$^{17}$  &20 &20 &20 &20 &20 & 20 &  &  &  &  &  & 120 \\
\hline
 \multicolumn{12}{|r|}{Total sum} & 4020 \\
\hline
\end{tabular}\label{tabfig5}}\end{table} 
 \begin{table}\caption{The fit parameters and coefficient of determination for
fitting the equation $A=\exp (d \times P + g)$ with the data presented in the Figure  \ref{fig5}. $E_0$ is the energy of showers.} 
\centering 
\resizebox{1\textwidth}{!}{\begin{tabular}{|c |c| c| c| c| c| c|}
\hline\hline                        
$E_0$ [eV] & $5\times10^{13}$ & $10^{14}$& $10^{15}$& $5\times10^{15}$& $10^{16}$& $10^{17}$ \\ [0.5ex]
\hline                  
{\it d} & $-3.37 \pm0.01$ &$ -3.46\pm 0.01$ &$-4.48\pm0.02$&$-4.63\pm0.02$&$-6.04\pm0.04$&$-7.67\pm0.06$   \\
{\it g} &$ 11.32\pm 0.05$&$ 12.23\pm0.06 $&$17.58\pm0.21$&$19.78\pm0.21$&$25.54\pm0.48$&$34.06\pm0.98 $   \\
$R^2$ & 0.998 & 0.994 &0.996&0.992&0.998&0.967    \\
\hline
\end{tabular}
\label{table3}}
\end{table}
 The fit parameters  $d$ and $g$, which are presented in Table \ref{table3}, are found to be energy dependent variables.  It is worth noting that a linear dependence of  $X_{max}$ on $\ln A$ and $\ln E$ is a known fact in extensive air showers. For example, Kampert and Unger \cite{Kampert12} have presented a linear equation, which reads as follows, after a parameter is renamed:
\begin{equation}
\langle X_{max}\rangle=f + D_p \ln (E/A), \label{equation3}
\end{equation}
in which the elongation rate for proton-initiated showers $D_p=\frac{d \langle X^p_{max}\rangle}{d\ln E}$ and $f$ are parameters that depend on the characteristics of hadronic interactions. By using equation \ref{equation3} to replace $X_{max}$ in equation \ref{equation1}, we obtain a quadratic relation between $P$ and $\ln A$:
\begin{equation}
\ln A=-\frac{a}{D_p} P^2  - \frac{b}{D_p} P + \frac{D_p \ln E +f-c}{D_p} , \label{equation4}
\end{equation}
in which $a$, $b$, and $c$ are parameters of equation \ref{equation1}. The last term in equation \ref{equation4} is clearly energy dependent . On the other hand, even though $D_p$ has almost  a constant value of 25 g/cm\textsuperscript{2} (see ref. \cite{Kampert12}), the coefficient of the first term in the linear model will bacome energy dependent  if we ignore the first term in equation \ref{equation4} to obtain the simpler linear model given in equation \ref{equation2}. Thus,  equation \ref{equation2} can be represented in
another form:
\begin{equation}
A=\exp (\alpha(E_0) \times P + \beta(E_0)), \label{equation5}
\end{equation}
in which $E_0$ is the energy of the shower. A
quadratic function of $\ln E_0$ is found to be suitable for $\alpha(E_0)$ and $\beta(E_0)$:
\begin{equation}
\alpha(E_0)=a_1 \ln E_0^2 + b_1  \ln E_0 + c_1 \label{equation6}
\end{equation}
\begin{equation}
\beta(E_0)=a_2  \ln E_0^2 + b_2   \ln E_0 + c_2 \label{equation7}
\end{equation}
in which $E_0$ is in TeV. The fit parameters for equations \ref{equation6} and \ref{equation7} are given in Table \ref{table4}.  Thus,
 equation \ref{equation5} is applicable for primary mass estimation if the shower
energy is known. Fortunately, previous research has shown that the shower energy
can be estimated from the surface density of Cherenkov radiation at a
fixed distance from the shower core. For example, Lubsandorzhiev \cite{Lubsandorzhiev08}
has presented the following relation between the shower energy and the
Cherenkov photon density at 175 m from the shower core:
\begin{equation}
E_0=400\times (Q_{175})^{0.95}, \label{equation8}
\end{equation}
where $Q_{175}$ is photon density in
cm\textsuperscript{-2}eV\textsuperscript{-1}, and
$E_0$ is the shower energy in TeV. This has been
utilized by Tunka Collaboration for the shower energy estimation.
 Therefore, we used equation \ref{equation8} to estimate the energies and then
equation \ref{equation5} to estimate the primary masses of another set of 1900 simulated
extensive air showers. The number of simulated showers for each type-energy is given in Table \ref{tabfig6}. It should be emphasized that the new set of
showers is different from that of showers used for the estimation of fit
parameters presented in Table \ref{table3} and Table \ref{table4}.
\begin{table}[ht]\caption{The fit parameters for
equations 6 and 7, evaluated from the data presented in Table \ref{table3}.  } 
\centering 
\begin{tabular}{|c| c| c| c| c|}
\hline\hline                        
Equation \ref{equation6} & $a_1$ & $b_1$ & $c_1$ & $R^2$\\ 
\hline                  
&$-0.062\pm0.005 $&$ 0.384\pm1.076 $&$ -3.941\pm12.97 $& 0.964  \\
\hline
Equation \ref{equation7} & $a_2$ & $b_2$ & $c_2$ & $R^2$\\ 
\hline                  
&$0.261\pm0.005 $&$ -1.035\pm1.076 $&$ 11.453\pm12.97 $& 0.984  \\
\hline
\end{tabular}
\label{table4}
\end{table}
\begin{table}[h]\caption{The number of simulated showers used  in Figure \ref{fig6} for the test of atomic mass estimation by equation \ref{equation5}.}
\centering
\resizebox{0.8\textwidth}{!}{\begin{tabular}{|c|c|c|c|c|c|c|c|c|c|c|c|c|}
\hline
 Energies[eV]  & p & He & C & O & Si & Fe & Mg & Al & S & K & Sc & Sum\\
\hline
10$^{14}$ & 50 &50 &50 & 50 & 50 & 50 &  &  & &  &  & 300  \\
\hline
5x10$^{14}$ &  & & 100 &100 &  & 100 & 100 & 100 & 100 &100  & 100 & 800 \\
\hline
10$^{15}$  & 50 &50 &50 &50 &50 & 50 &  &  &  &  &  & 300 \\
\hline
2x10$^{15}$  &  &  & 100 &100 & & 100 &  &  &  & 100 & 100 & 500 \\
\hline
 \multicolumn{12}{|r|}{Total sum} & 1900 \\
\hline
\end{tabular}\label{tabfig6}}\end{table} 
The histogram of deviation of the estimated primary masses from the actual
masses is depicted in Figure \ref{fig6}. The mean of the deviation distribution is
close to zero. This means the method can discriminate the average
primary mass of cosmic rays. However, the large value of the RMS
implies a statistical error of 19 a.m.u. for a single event mass
estimation.
  \section{Impact of the hadronic interaction models on the results}
\begin{figure}
\includegraphics[width=0.6\textwidth]{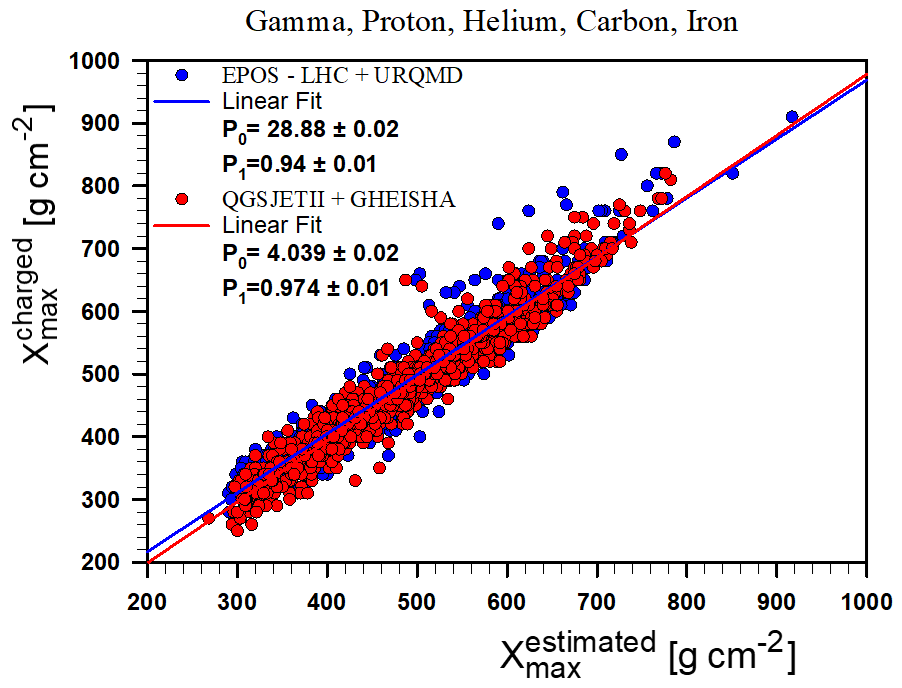}
\centering\caption{   A comparison of $X_{max}^{estimated}$ obtained by using equation \ref{equation1} with $X_{max}^{charged}$ for showers initiated by gamma ray, proton, helium, carbon and iron primaries of $10^{14}$, $10^{15}$ and $10^{16}$ eV energies. The blue dots are for showers generated with EPOS-LHC+UrQMD models, and the red dots are for those  generated by QGSJETII+GHEISHA models. Each population has been fit to a linear model $X_{max}^{charged}=p_0+p_1 X_{max}^{estimated}$. The corresponding fit parameters are given in the legend. The number of showers used for each type-energy-model can be found in Tables \ref{tabfig2} and \ref{tabfig7}.}\label{fig7}
 \end{figure}

\begin{figure}
\centering\includegraphics[width=0.6\textwidth]{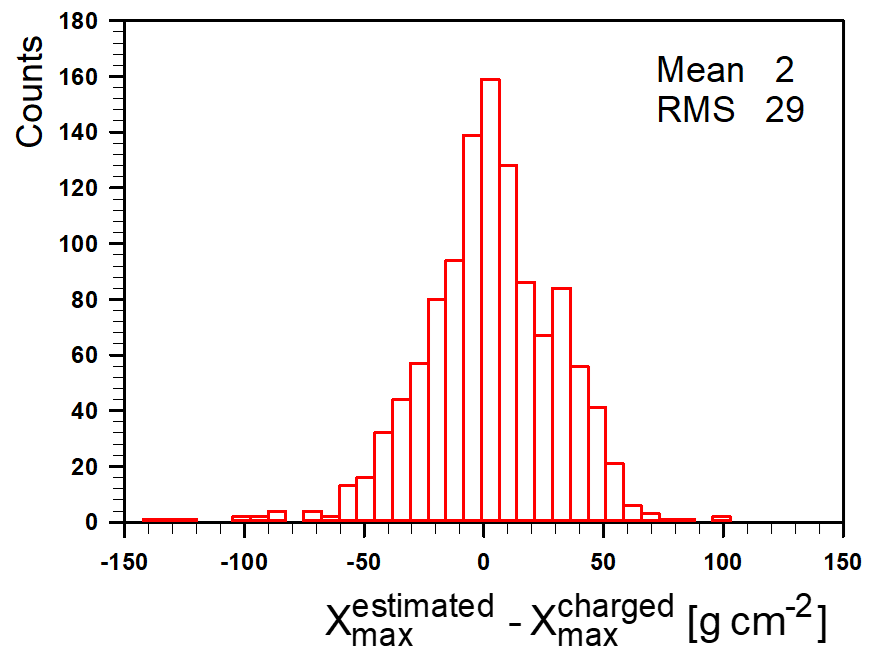}
\caption{  A histogram of difference of $X_{max}^{estimated}$ with $X_{max}^{charged}$ for showers initiated by gamma ray, proton, helium, carbon and iron primaries of $10^{14}$, $10^{15}$ and $10^{16}$ eV energies generated with EPOS-LHC+UrQMD hadronic models.  The number of showers used for each type-energy can be found in Table \ref{tabfig7}.}\label{fig8}
 \end{figure}

\begin{figure}
\centering\includegraphics[width=0.6\textwidth]{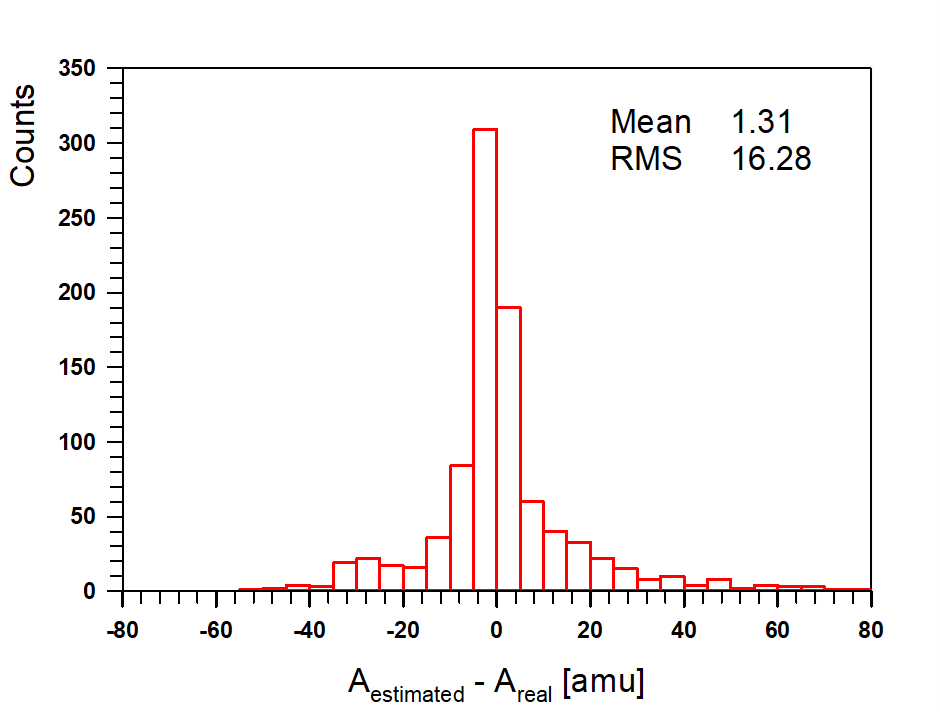}
\caption{  The distribution of differences between the actual masses of
the primary particles and the estimated masses obtained by application
of equation \ref{equation2}. The results are based on event-by-event evaluation for
920 simulated extensive air showers initiated by proton, helium, carbon and iron primaries of $10^{14}$, $10^{15}$ and $10^{16}$ eV energies, which were generated with EPOS-LHC+UrQMD hadronic models.  The number of showers used for each type-energy can be found in Table \ref{tabfig7}.}\label{fig9}
 \end{figure}

 All the analyses presented in previous sections were based on showers simulated using QGSJETII and GHEISHA hadronic interaction models. In order to test the impact of hadronic interaction models on the proposed method of $X_{max}$ estimation, we simulated a set of $10^{14}$, $10^{15}$ and $10^{16}$eV showers initiated by five different primary masses  with EPOS-LHC model for high energy, and UrQMD model for low energy hadronic interactions.  The numbers of simulated showers are given in Table \ref{tabfig7}. 
\begin{table}[h]\caption{Number of simulated showers with EPOS-LHC+UrQMD hadronic models.}
\centering
\resizebox{0.5\textwidth}{!}{\begin{tabular}{|c|c|c|c|}
\hline
 & \multicolumn{3}{c|}{No. of showers} \\
\hline
Type$\backslash$ Energy & 10$^{14}$eV  & 10$^{15}$eV & 10$^{16}$eV  \\
\hline
$\gamma$ & 100 & 100 & 30 \\
\hline
 \it{p}  & 100 & 100 & 30 \\
\hline
 He  & 100 & 100 & 30  \\
\hline
C & 100 & 100 & 30 \\
\hline
Fe & 100 & 100 & 30\\
\hline
\end{tabular}\label{tabfig7}}\end{table}
 The steepness parameter of each shower was then used in equation \ref{equation1} to estimate its  $X_{max}$. It should be noted that the parameters of equation \ref{equation1}  are obtained by using showers generated through QGSJETII+GHISHA models. A comparison of estimated depths with those obtained from longitudinal distribution of charged particles of the showers, together with the similar results for showers of the same type-energy, generated by QGSJETII+GHEISHA models, is presented in Figure \ref{fig7}. The average values of estimated and actual $X_{max}$ for the EPOS-UrQMD showers are given in Table \ref{table5}. A histogram of  $X_{max}^{estimated} - X_{max}^{charged}$ is shown in Figure \ref{fig8}. Despite differences in the slopes and offsets of the linear fits in the two sets of hadronic models presented in Figure \ref{fig7}, the statistical error of the estimated $X_{max}$ of EPOS-UrQMD showers, which is inferred from Figure \ref{fig8} to be about 30 g/cm\textsuperscript{2}, is lower than the difference between the average $X_{max}$s of low mass and high mass cosmic rays. 
 The application of equation \ref{equation2} has also been tested on the EPOS+UrQMD showers initiated by proton, helium, carbon and iron. The histogram of deviation of estimated mass from actual mass is shown in Figure \ref{fig9}. The RMS in the estimated mass deviations is comparable to the one obtained for QGSJETII+GHEISHA showers (see Figure \ref{fig6}).   Although different hadronic models can generate slightly different  $X_{max}$ for showers of the same type-energy, it seems that the relation between the $X_{max}$ and the lateral distribution of Cherenkov radiation is not seriously affected.   In fact, hadronic interaction models are
expected to diverge at  energies higher than $10^{17}$ eV  where no accelerator data are available. As we did not have enough  computation time to simulate higher energy showers, the presented test is
limited to  $[10^{14},10^{16}]$ eV energy range. Further tests should be
done for higher energies.  The results of the maximum depths  for showers simulated through EPOS-LHC and QGSJETII hadronic models in energies above $10^{17}$ eV reported by  Yushkov et al. show that the difference between average $X_{max}$ in the two models is lower than 30  g/cm\textsuperscript{2} \cite{Yushkov19}.  
\begin{table}[ht] \caption{ The average values for the steepness of the lateral
distribution of Cherenkov radiation, \emph{P}, the maximum depths
obtained from longitudinal distribution of charged particles, $X_{max}^{charged}$ , the
maximum depths estimated by using equation \ref{equation1}, $X_{max}^{estimated}$, and the relative
difference of the two maxima, $\delta=\frac{\vert X_{max}^{charged} - X_{max}^{estimated}\vert}{X_{max}^{charged}}$, for the showers simulated with EPOS+UrQMD hadronic interaction models. The number of simulated showers used for calculation of these data is given in Table \ref{tabfig7}. }
\centering
\resizebox{\textwidth}{!}{\begin{tabular}{|c|c|c|c|c|c|}
\hline
 Primary Particle & Primary Energy [eV] & $P=Q(100)/Q(200)$  & $X_{max}^{charged}$ [g/cm$^2$] & $X_{max}^{estimated}$ [g/cm$^2$] & $\delta$ \\
\hline
$\gamma$-ray  & $10^{14}$  & $3.87 \pm0 .04$ & $510 \pm 6.6$ & $544 \pm 6.8$ &     0.067 \\              
        & $10^{15}$         & $4.28 \pm 0.03$ & $579 \pm 5.8$ & $614 \pm 5.9$ &     0.061  \\  
        & $10^{16}$         & $4.63 \pm 0.04$ & $660 \pm 9.5$ & $678 \pm 6.9$ &     0.027  \\  
\hline           
\it{p}  & $10^{14}$  & $3.41 \pm 0.05$ & $484 \pm 10$ & $472 \pm 7.6$ &     0.025 \\              
        & $10^{15}$         & $3.87 \pm 0.05$ & $543 \pm 9.8$ & $546 \pm 8.1$ &     0.004  \\             
        & $10^{16}$         & $4.14 \pm 0.04$ & $590 \pm 10.6$ & $590 \pm 7.5$ &     0.  \\             
\hline
He      & $10^{14}$  & $3.14 \pm 0.03$ & $437 \pm 6.1$ & $430 \pm 4.9$ &     0.016 \\              
        & $10^{15}$         & $3.48 \pm 0.02$ & $531 \pm 5.8$ & $528 \pm 4.9$ &     0.006  \\             
        & $10^{16}$         & $4.06 \pm 0.05$ & $580 \pm 11.1$ & $575 \pm 8.2$ &     0.009  \\             
\hline
C       & $10^{14}$  & $2.86 \pm 0.02$ & $398 \pm 5.5$ & $389 \pm 3.4$ &     0.023 \\              
        & $10^{15}$         & $3.48 \pm 0.02$ & $479 \pm 4.5$ & $479 \pm 3.3$ &     0.  \\             
        & $10^{16}$         & $3.87 \pm 0.04$ & $546 \pm 8.2$ & $543 \pm 6.4$ &     0.005  \\             
\hline
Fe      & $10^{14}$  & $2.35 \pm 0.01$ & $338 \pm 2.8$ & $321 \pm 1.7$ &     0.050 \\              
        & $10^{15}$         & $3.08 \pm 0.01$ & $423 \pm 2.6$ & $420 \pm 1.9$ &     0.007  \\      
        & $10^{16}$         & $3.62 \pm 0.02$ & $501 \pm 4.6$ & $501 \pm 3.3$ &     0.  \\      
        \hline       
\multicolumn{5}{|r|}{ $\langle\delta\rangle=$ } &0.020\\
\hline
\end{tabular}\label{table5}}\end{table}

\section{Conclusion}
One of the main goals of observing extensive air showers is the estimation
of the mass of their primary cosmic rays. The mass is usually estimated
by measuring the shower maximum depth with air fluorescence detectors
\cite{Abbasi18}, or by measuring the muon to electron population ratio with an
array of particle detectors \cite{Purmohammad13}. Since the Cherenkov wave front of
an extensive air shower is mainly generated by high energy electrons,
the depth of maximum of a shower is the origin of the most of the
radiation. The lateral distribution of the Cherenkov radiation on the
ground is, then, affected by variation of the shower maximum depth.
Although the details of Cherenkov lateral distribution depend on the
variation of the energy distribution of the secondary electrons, and
structure of the atmosphere, the Monte Carlo simulation technique
provides the appropriate means to investigate the phenomena. Researchers
in Tunka experiment have used the slope parameter of the lateral
distribution of Cherenkov radiation in extensive air showers, for
the estimation of the height of the shower maximum. The estimated height, 
then, has to be converted to atmospheric depth in order to be used for
the primary mass estimation. Here, we indicated that the maximum depth
could be immediately related to the slope parameter, with no need to the
intermediate step of maximum height estimation. The maximum depths
estimated with the new method are more accurate. This, in turn, can
improve the accuracy of the cosmic ray mass composition estimated by
Cherenkov wave front experiments, like Tunka-133 array \cite{Prosin14}. We have
also tested a new relation between the primary mass and the slope
parameter. The method has been tested for the estimation of the primary
masses from the measured slope of the lateral Cherenkov radiation of
simulated extensive air showers. Although the idea is useful for average
mass estimation, the statistical errors are large. This prevents
the application of the method for an accurate estimation of the primary mass
of a single shower. It should be mentioned that the detector array
response has not been taken into account in our analysis. In reality,
the estimated position of a shower core and the photon densities in an
array have large instrumental errors. These errors can reduce the
accuracy of the results. It is worth noting that due to limited computation
time, our simulated data were restricted to the vertical showers. Both
shower zenith angle and detector responses can affect the accuracy of
the results. Further investigations should take account of a specific detector
array response and dependence of Cherenkov lateral distribution on the
shower zenith angle.

\section*{Acknowledgements}
Most of the simulation in this work has been performed with the HPC
supercomputer at Imam Khomeini International University (IKIU). We are grateful
to the manager and the staff of HPC facility of IKIU for their support. 

\bibliography{mybibfileR2}
\end{document}